\documentclass[showpacs,prl,aps,superscriptaddress,twocolumn,floatfix]{revtex4}
\usepackage[T1]{fontenc}
\usepackage{graphicx}
\usepackage{amsmath,float}
\usepackage{amsfonts}
\usepackage{amsthm}
\usepackage{amssymb}
\usepackage[a4paper,left=1cm,right=1cm,top=2cm,bottom=2cm]{geometry}
\usepackage{bbm}

\date{\today}
\begin{document}
\title{Magnetic field induced localization in 2D topological insulators}

\author{Pierre Delplace}
\author{Jian Li}
\author{Markus B\"uttiker}
\affiliation{D\'epartement de Physique Th\'eorique, Universit\'e de Gen\`eve,
CH-1211 Gen\`eve, Switzerland}


\begin{abstract}
Localization of the helical edge states in quantum spin Hall insulators
requires breaking time reversal invariance. In experiments this is naturally
implemented by applying a weak magnetic field $B$. We propse a model based on
scattering theory that describes the localization of helical edge states
due to coupling to random magnetic fluxes. We find that the localization length
is proportional to $B^{-2}$ when $B$ is small, and saturates to a constant
when $B$ is sufficiently large. We estimate especially the localization
length for the HgTe/CdTe quantum wells with known experimental parameters.
\end{abstract}

\maketitle
 
The prediction and discovery of quantum spin Hall insulators (QSHIs)
\cite{Kane:2005,kane_z2_2005,bernevig06,konig07} has opened a door
to an unexpected category of topological phases in condensed matter
\cite{kitaev_periodic_2009,schnyder_classification_2009,hasan_colloquium_2010,
Qi:2011}, and revealed a new route to investigations of edge/boundary-state
physics \cite{roth_nonlocal_2009, buttiker_edge-state_2009, brune12}. Although
the prototypes of QSHIs \cite{Kane:2005,bernevig06} are
mainly based on two copies of quantum Hall insulators, which have been
investigated for more than three decades \cite{klitzing80,Haldane:1988}, it was
soon realized that the fundamental importance of time reversal invariance (TRI)
distinguishes the two systems in a profound way \cite{kane_z2_2005}. Indeed,
QSHIs, unlike the $\mathbb{Z}$-classified quantum Hall
insulators \cite{thouless_quantized_1982}, belong to a class of two-dimensional
time-reversal-invariant $\mathbb{Z}_2$ topological insulators
\cite{kane_z2_2005}. The defining feature of QSHIs, as its name suggests, is a
pair of helical edge states that persist in the bulk insulating gap of the
system \cite{Kane:2005,kane_z2_2005,bernevig06,konig07,roth_nonlocal_2009}.

The topological power of QSHIs lies precisely in the
robustness of the helical edge states against generic perturbations due to
unavoidable disorder in every experimental setup, unless TRI is broken. In the
presence of both TRI breaking and disorder, the helical edge states will be
localized, and the general framework of Anderson's
localization theory applies \cite{anderson58}. Nevertheless, the localization of the
helical edge states distinguishes itself from conventional one-dimensional
localization when the focus is placed on the crucial role TRI plays in the
problem. This point becomes especially relevant as TRI can be broken
continuously, for instance, by turning on a magnetic field gradually.
 Indeed, the sensibility of transport though helical edge states to
weak magnetic field has been demonstrated experimentally in the measurement of
magneto-conductance in topologically nontrivial HgTe/CdTe quantum wells
\cite{konig07,konig08}. Related theoretical analyses have been carried out that
include the interplay between TRI-breaking and disorder, but mainly consider
magnetic impurities \cite{tanaka11,hattori11}, or bulk random potential combined
with magnetic field \cite{maciejko10}. A transparent edge theory that focuses on
the magnetic-field-dependent localization of the helical edge states, however,
is still missing.

\begin{figure}
\centering
\includegraphics[width=0.4\textwidth]{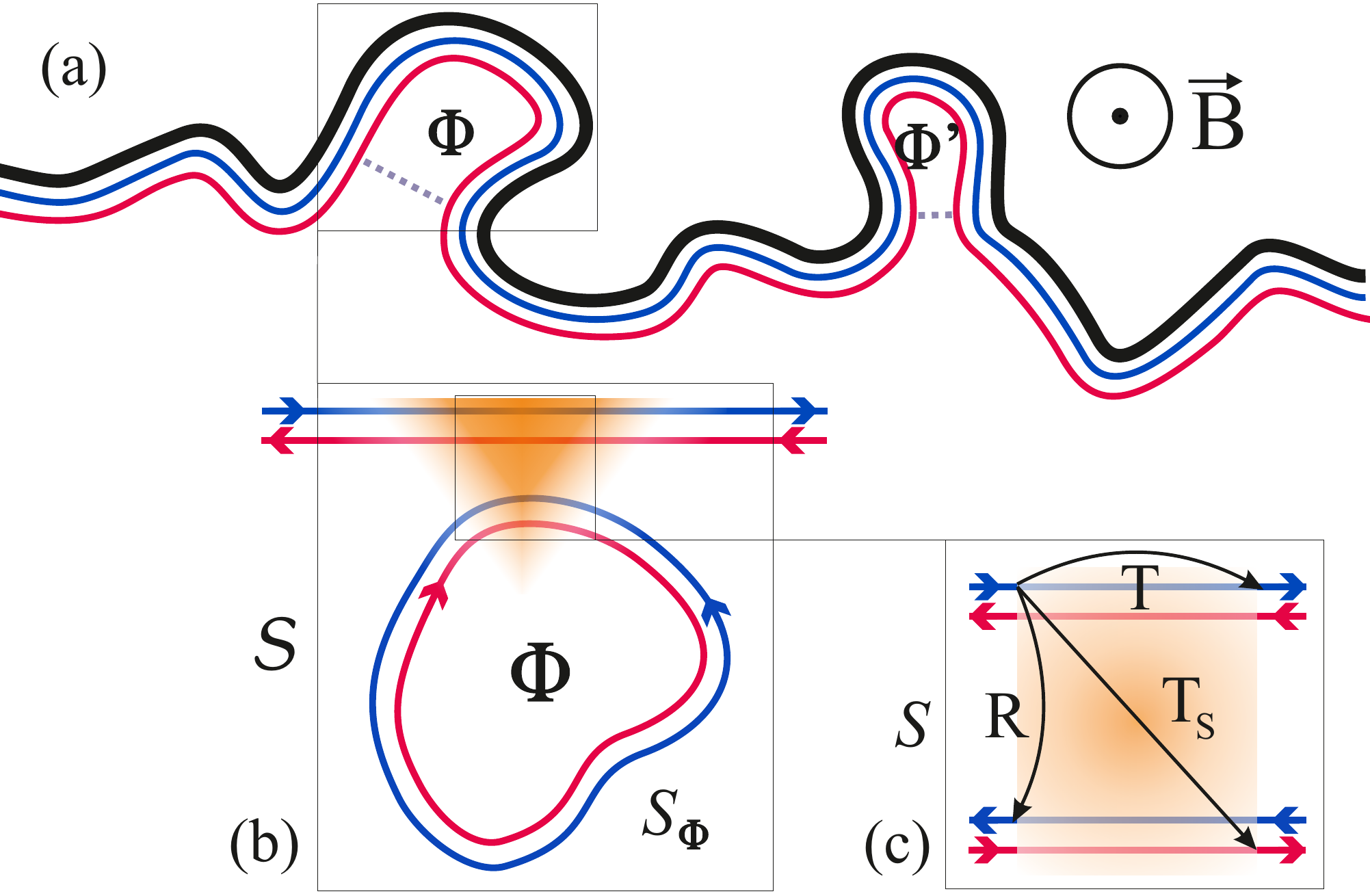}
\caption{(a) Helical edge states in a disordered QSHI in a uniform
magnetic field. Occasional occurrences of constrictions along the edge lead to
Fabry-Perot-type loops where Aharonov-Bohm phases due to magnetic fluxes can
accumulate. (b) The scattering of the helical edges by one of these loops,
described by a scattering matrix ${\cal S}$, can be divided into two parts: the
scattering between two pairs of helical edge states ($S$),
and the propagation of one of these pairs around the loop ($S_\Phi$).
(c) Three types of scattering probabilities, $T$, $R$ and $T_s$, that are
relevant to the scattering between two pairs of helical edge states.}
\label{fig:edge}
\end{figure}

In this paper we propose a model that explicitly addresses the question on
how the localization of helical edge states occurs as a weak magnetic field is
gradually turned on. Our model is based on the scattering theory of edge
states in the presence of generic edge disorder. In particular we consider the
existence of alternative paths for the edge states due to, e.g., constrictions
formed at the rough edges of realistic samples (see Fig. \ref{fig:edge}),
which further allows for loops of the helical edge states. The magnetic field
penetrating though these loops results in broken TRI that is experienced by
the helical edge states in the form of finite random fluxes. We show that these
random fluxes necessarily lead to localization of the helical edge states with
universal behaviors in two regimes: immediately after the magnetic field is
turned on, the localization length becomes finite and decreases as $B^{-2}$;
when the standard deviation of the random fluxes (proportional to $B$) is
comparable to or larger than one magnetic flux quantum, the localization length
saturates to a constant. In-between these two regimes, damped oscillations of
the localization length may arise depending on the distribution of the random
fluxes. Our results provide a clear illustration of the
symmetry-breaking-induced localization in QSHIs.

We start to introduce our model by considering one of its possible realizations,
depicted in Fig. \ref{fig:edge}. The edge roughness of a realistic
QSHI sample may lead to occasional constrictions
(at one edge) where the helical edge states can either tunnel across or pass
around. As a consequence, loops can form and attach to the propagating
path of the helical edge states. When a perpendicular magnetic field is applied
to the sample \footnote{We focus on the weak magnetic field regime, such that
the Zeeman effect is negligible. For the same reason, we also can ignore the effect of magnetic field on
the bulk bands.}, each of these loops acts as a
\textit{magnetic flux impurity}, to be distinguished from usual magnetic
impurities. Individually such an impurity works like a Fabry-Perot
scatterer (see Fig. \ref{fig:edge}b), where the scattering probability
amplitudes depend on the Aharonov-Bohm (AB) phase, owing to the magnetic flux,
acquired by electrons circling around the loop. The collective action of a
random distribution of these scatterers causes localization of the helical edge
states with an explicit dependence on the magnetic field. The main part of this
problem can be tackled consistently by scattering theory, as we now show.

To analyze the scattering of the helical edge states by a single magnetic flux
impurity (see Fig. \ref{fig:edge}), we divide the full scattering process into
two effective parts: the local scattering between two pairs of helical states,
and the free propagation of one pair of helical states that closes the loops.
For simplicity, we assume that the first part does not depend on magnetic field,
hence respects \textit{local} TRI, while the second part contains the entire
information about the magnetic flux by means of AB phases that enter the
propagators.

Owing to the \textit{local} TRI, the scattering between two
pairs of helical states, described by a $4\times 4$ scattering matrix $S$, has
the following constraint:
\begin{equation}\label{eq:trs}
  S = \Theta \ S^\dagger\ \Theta^{-1},
\end{equation}
where ${\Theta}$ is the time-reversal operator. We choose a specific basis
ordered as $(R_1,L_1,L_2,R_2)$, where $R_i$ ($L_i$) stands for the right (left)
mover of the $i$-th  Kramers pair ($i=1,2$), such that the time-reversal
operator reads
\begin{equation}\label{eq:tro}
  \Theta =
  \begin{pmatrix}
    -i\sigma_y \kappa & 0 \\
    0 & -i\sigma_y \kappa \\
  \end{pmatrix}
\end{equation}
with $\sigma_y$ the Pauli matrix and $\kappa$ the complex-conjugate operator. %
Consequently, the scattering matrix $S$ satisfying Eq. \eqref{eq:trs}
necessarily has the following form:
 \begin{equation}\label{eq:s4}
S=\left(
\begin{array}{cccc}
t_1         &     0             &    r'          &    s' \\
0           &     t_1           &    -s          &    r  \\
r           &     -s'           &    t_2         &    0  \\
s           &     r'            &    0           &    t_2
\end{array}
\right)\,.
\end{equation}
Here $t_i$ stands for the direct transmission for Kramers pair $i$; $r$ ($r'$)
stands for the reflection from a right (left) mover to a left (right) mover; $s$
and $s'$ represent the transmission by switching to another Kramers pair.
Importantly, zeros in $S$ signify the absence of direct back-scattering within
one Kramers pair due to TRI. Taking into account the
unitarity of the scattering matrix, the parametrization of $S$ can be further
simplified as (up to an unimportant global phase factor): $t_1 = -t_2^* = t$,
$r' = r^*$, $s' = s^*$, and $T+R+T_s=1$, where $T=|t|^2$, $R=|r|^2$ and
$T_s=|s|^2$.

The free propagation of Kramers pair 2 leads to a $2\times 2$ scattering matrix
$S_\Phi$ that is diagonal in the basis $(L_2,R_2)$, given by
\begin{equation}\label{eq:sphi}
S_\Phi =\left(
 \begin{array}{cc}
 e^{i(\varphi+\phi)} &  0\\
0           &  e^{i(\varphi-\phi)}
 \end{array}
\right)
\end{equation}
with $\varphi$ the dynamical phase and $\phi$ the AB phase (equal to the total
flux enclosed by the loop in units of $h/2\pi e$).

Combining the two parts above, we find the final scattering
matrix ${\cal S}$ for Kramers pair 1, in the basis $(R_1,L_1)$, to be
\begin{align}\label{eq:cals}
{\cal S}
% &= S_{1,1} + \sum_{n=0}^{\infty} S_{1,2}  (S_\Phi S_{2,2})^n   S_\Phi S_{2,1}
= \begin{pmatrix}
    |t| + RZ_+ + T_sZ_- & -(rs)^*\Delta Z \\
    -rs\Delta Z & |t| + RZ_- + T_sZ_+ \\
  \end{pmatrix},
\end{align}
\begin{gather}
\text{where}\qquad  Z_{\pm} = \frac{e^{i(\varphi\pm\phi)}}{1+|t|e^{i(\varphi\pm\phi)}}\,, \\
  \Delta Z = Z_+-Z_- = \frac{i}{|t|}\,\frac{\sin\phi}{\cos\phi+\cosh(\ln |t|
+i\varphi)}\,,
\end{gather}
and the phase of $t^*$ has been absorbed into the dynamical phase $\varphi$. The
back-scattering probability can be obtained immediately:
\begin{align}\label{eq:calr}
{\cal R} = \frac{RT_s}{T}\,\frac{\sin^2\phi}{\bigl|\cos\phi+\cosh(\ln |t|
+i\varphi)\bigr|^2}\,.
\end{align}
Evidently, for the helical edge states to be back-scattered with finite
probability, two conditions must be fulfilled. First, it is necessary that both
$R$ and $T_s$ are finite. If one of these two tunneling probabilities is zero,
the system essentially reduces to two decoupled copies of quantum Hall edge
states, and back-scattering is known to be prohibited for either copy \cite{buttiker88}.
 Only when both tunneling processes ($R$ and $T_s$) are allowed, the
system belongs truly to the $\mathbb{Z}_2$-classified symmetry class where TRI
plays a central role. It follows that the second necessary condition for
back-scattering to occur is to break the \textit{global} TRI by having $\phi \ne
0\mod\pi$. The cooperation of these two conditions clearly illustrates the
underlying protection mechanism, from the scattering point of view, for the
helical edge states of QSHIs.

At disordered sample edges, the magnetic flux impurities will occur randomly,
and the helical edge states will eventually be localized as a consequence of
finite back-scattering probabilities for individual scatterers. Here, we assume
not only that the variables (including phases and scattering amplitudes) for
each individual scatterer are random, but also that different scatterers are
completely independent such that the relative scattering phases for two
consecutive scatterers are uniformly distributed. The localization length of the
helical edge states in this scenario can be extracted from the appropriate
scaling variable $\ln{\cal T}$, where ${\cal T} = 1-{\cal R}$ is the total
transmission probability for the effective one-dimensional system \cite{ATAF80}.

The total transmission probability ${\cal T}_N$ through $N$ scatterers can be
calculated by multiplying the transfer matrices that relate the scattering
amplitudes on the right side of each individual scatterer (labeled $i$,
$i=1,...,N$) to those on the left. A general transfer matrix reads
\begin{equation}
\label{eq:Mi}
M_i = \frac{1}{\tau_i}\,\left(
\begin{array}{cc}
\lambda_i   &   \rho_i   \\
 \rho^*_i   &   \lambda^*_i
\end{array}
\right),
\end{equation}
where $\tau_i$ corresponds to the transmission amplitudes (diagonal entries in
${\cal S}$) for the $i$-th scatterer, $\rho_i$ corresponds to the reflection
amplitudes (off-diagonal entries in ${\cal S}$), and $\lambda_i$ is a phase
factor that is independent for each scatterer. Note that the dynamical phase for
the free propagation of the helical states in-between two consecutive scatterers
($i$ and $i+1$, say) can be obviously incorporated into the above transfer
matrix while preserving its general form. Note also that multiplications of the
transfer matrices preserve the general form as well. We will put an overhead
tilde to distinguish the amplitudes involving $i$ consecutive scatterers from
those involving only a single ($i$-th) scatterer. Then ${\cal T}_N$ is given by
\begin{equation}
\label{eq:TN}
{\cal T}_N = |\tilde{\tau}_{N}|^2 =
\frac{|\tilde{\tau}_{N-1}|^2|\tau_N|^2}{\bigl|1+|\tilde{\rho}_{N-1}\rho_N|e^{
i\theta}\bigr|^2},
\end{equation}
with $e^{i\theta} =
\tilde{\lambda}_{N-1}\lambda_N\tilde{\rho}_{N-1}\rho_N^*/|\tilde{\rho}_{N-1}
\rho_N|$. Our assumption of independent scatterers implies that $\theta$ is
uniformly distributed in $[0,2\pi)$. It follows that the mean of the scaling
variable $\langle\ln{\cal T}\rangle$ becomes simply additive \cite{ATAF80}
\begin{equation}
\label{eq:Tadd}
\langle\ln{\cal T}_N\rangle = \langle\ln{\cal T}_{N-1}\rangle +
\langle\ln|\tau_N|^2\rangle,
\end{equation}
where $\langle\ln\bigl|1+|\tilde{\rho}_{N-1}\rho_N|e^{i\theta}\bigr|^2\rangle$
has vanished after averaging over $\theta$.

The inverse localization length $\gamma = 1/\ell_\text{loc}$ is defined in terms
of the scaling variable as: \cite{ATAF80,pendry,houches09}
\begin{equation}
\label{eq:gamma}
  \gamma = - \lim_{N \rightarrow \infty} \frac{n}{N} \left\langle \ln{\cal
T}_N\right\rangle = -n{\langle\ln(1-{\cal R})\rangle},
\end{equation}
where $n$ is the linear density of the scatterers. The final average
$\langle\ln(1-{\cal R})\rangle$ is over certain distributions of independent
variables including $\varphi$, $\phi$ and scattering amplitudes that appear in
Eq. \eqref{eq:calr} for a \textit{single} scatterer. We are interested in the
weak back-scattering case for each individual scatterer (${\cal R} \ll 1$), thus
\begin{equation}
\label{eq:gamma2}
  \gamma \simeq n{\langle{\cal R}\rangle}.
\end{equation}
The average in terms of the arbitrary dynamical phase $\varphi$
can be carried out exactly, and yields
\begin{equation}
 \label{eq:gamma3}
 \gamma  =
n\left\langle\frac{RT_s}{R+T_s}\,\frac{1+T}{T}\,\frac{\sin^2\phi}{\sin^2\phi +
\frac{(1-T)^2}{4T}}\right\rangle.
\end{equation}
By further using the fact that $\phi = BA(2\pi e/h)$, where the magnetic field
$B$ is taken to be uniform and $A$ is the area enclosed by the helical loops, we
will only need to average over distributions of the scattering amplitudes and
the area $A$ in order to estimate the localization length $\ell_\text{loc} =
1/\gamma$. One immediate consequence of Eq. \eqref{eq:gamma3} is that the
localization length is magnetic field symmetric, which is certainly expected.

An important regime that we are particularly interested in is the weak magnetic
field regime, where $B\bar{A} \ll h/e$ with $\bar{A}$ the mean of $A$. In this
regime Eq. \eqref{eq:gamma3} becomes (assuming $T$ is not too close to $1$)
\begin{gather}
 \gamma  = \alpha B^2 \label{eq:gamma_Bsmall}\\
 \text{with}\qquad \alpha = 4n \Bigl(\frac{2\pi e}{h}\Bigr)^2
\Bigl\langle\frac{RT_s}{(R+T_s)^3}(1+T)\Bigr\rangle \langle A^2 \rangle\,.
\label{eq:alpha}
\end{gather}
Manifestly $\alpha$ is a constant factor for given distributions of scattering
amplitudes and $A$. The $B^2$-dependence of the inverse localization
length here is a \textit{universal} result of our model in the sense that it
does not depend on the specific distributions of variables for individual
scatterers. It implies that the localization length of the helical edge states
is finite at weak magnetic field and diverges only as $1/B^2$ when the magnetic
field is vanishing. Furthermore, the low-temperature magneto-conductance of a
QSHI should also vary as $B^2$ in the weak magnetic field limit,
that is, $\delta G(B) = G(0) - G(B) \propto B^2$. This contrasts our result with
the linear magneto-conductance behavior previously found on a lattice model
with bulk impurity potentials \cite{maciejko10}.

Another interesting regime is when the magnetic field is strong enough such that
both $B\bar{A}$ and $B\sigma_A$ (with $\sigma_A \equiv \sqrt{\langle
(A-\bar{A})^2 \rangle}$ being the standard deviation) are comparable or larger
than a flux quantum $h/e$. In this regime we can approximate the average in
terms of $A$ as an average over a uniform distribution of $\phi$, which yields
\begin{equation}
 \gamma_{\text{sat}}  = 2n\left\langle\frac{RT_s}{R+T_s}\right\rangle\,.
\label{eq:gamma_Blarge}
\end{equation}
This especially simple result again shows a \textit{universal} behavior in our
model--the inverse localization length saturates at relatively strong magnetic
field irrespective of the specific distributions of scattering variables. However
the actual value of $\gamma_{\text{sat}}$ certainly depends on the distributions of $R$ and $T_s$.
 Moreover, the above formula re-emphasizes the importance of allowing both tunneling processes
represented by $R$ and $T_s$ to evoke a true protection mechanism due to TRI.

\begin{figure}
\centering
\includegraphics[width=0.4\textwidth]{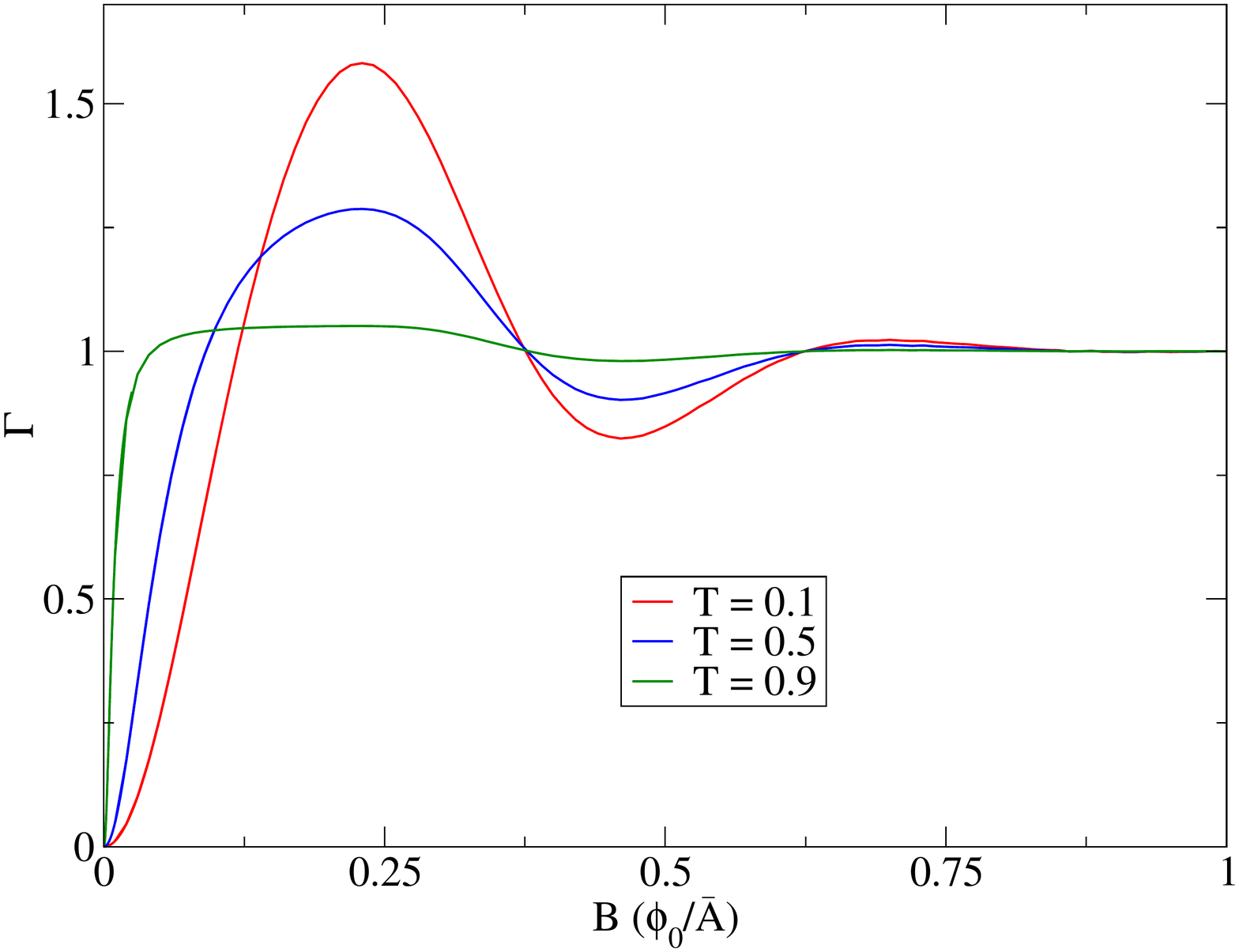}
\caption{$\Gamma(B)$ for different values of $T$,
where $A$ is assumed to take a normal distribution with $\sigma_A/\bar{A}=0.3$.}
\label{fig:Gamma1}
\end{figure}

In-between the two regimes discussed above, we need to consider the specific
distributions of the variables in Eq. \eqref{eq:gamma3}.
 Let us first focus on the behavior of $\gamma$ by assuming that $A$ has a
Gaussian distribution characterized by the mean $\bar{A}$ and the standard deviation
$\sigma_A$.  It is instructive in this case to look at the function $\Gamma(B)
\equiv [(1+T)/2T] \left\langle \sin^2\phi / [\sin^2\phi + (1-T)^2/4T]
\right\rangle_{A}$ with the average only taken in terms of $A$. $\Gamma(B)$ has
been defined such that it saturates to the value $1$ at sufficiently strong
magnetic field. In Fig. \ref{fig:Gamma1} we plot the numerically obtained
$\Gamma(B)$ for various $T$ and fixed $\bar{A}$ and $\sigma_A$. Right after the
magnetic field is turned on, $\Gamma(B)$ shows a quadratic increase irrespective
of the assumed $T$ or the distribution of $A$. Before $\Gamma(B)$ saturates, it
undergoes damped oscillations when $\sigma_A/\bar{A}$ is small. These
oscillations are due to the collective AB effect for the helical loops in our
model: the loops enclosing similar area lead to AB oscillations of similar
period; they contribute coherently to the back-scattering of the helical edge
states; the magnetic field dependence of the total transmission is thus shaped
by the AB effect at individual scatterers when $B\sigma_A$ is significantly
smaller than $\phi_0 = h/e$. The period of the oscillations is roughly
$\phi_0/2\bar{A}$, where the factor $1/2$ is obviously a consequence of $\Gamma$
(and hence $\gamma$) only depending on $\sin^2\phi$. \footnote{Note that the
period for the back-scattering probability ${\cal R}$ at a single scatterer is
$\phi_0/\bar{A}$ without the factor $1/2$.} The overall amplitude of the
oscillations is suppressed for large $T$ and enhanced for small $T$. The reason
is intuitively clear: the more the helical edge states are scattered into the
loops, the more pronounced the resulting AB effect.

\begin{figure}
\centering
\includegraphics[width=0.3\textwidth]{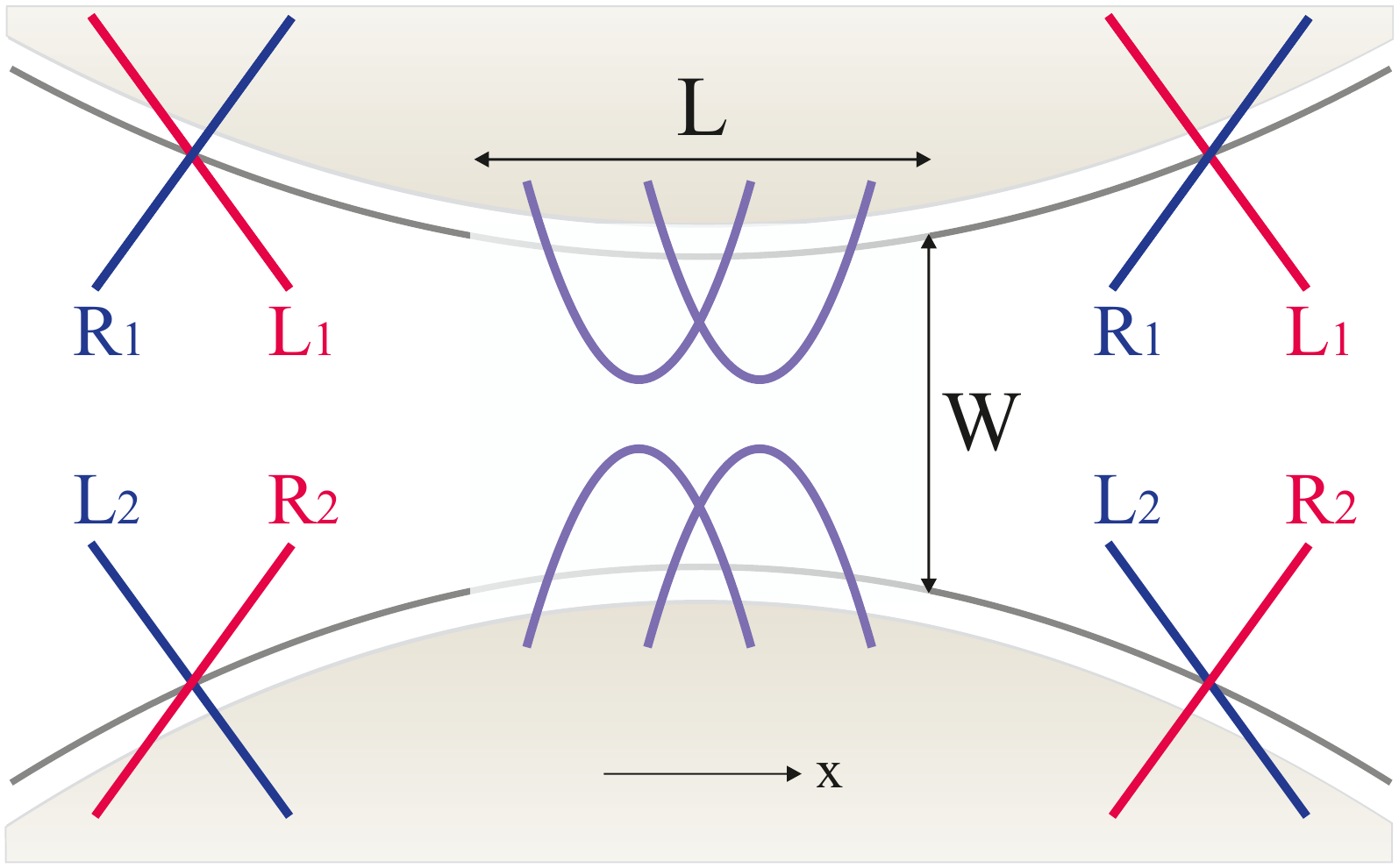}
\caption{A schematic view of a constriction where two pairs of helical edge
states (indicated by the linear bands) are coupled (indicated by the mixed
bands) in a region of length $L$ and separation $W$.
%The in-coming right and left movers can either be scattered onto the other edge
%or be transmitted through the constriction but cannot
%be back-scattered along the same edge because of the TRI.
}
\label{fig:constriction}
\end{figure}

Now we address the issue of the scattering amplitudes which have only been
assumed to be phenomenological parameters in the scattering matrix $S$ so far.
To this end we investigate a constriction depicted in Fig.
\ref{fig:constriction}, which is described by the following effective
Hamiltonian:
\begin{equation}\label{eq:Ham1D}
  {\cal H} =
  \begin{pmatrix}
    \hbar v_F \hat{k}_x & 0 & m(x) & \delta(x) \\
    0 & -\hbar v_F \hat{k}_x & -\delta(x) & m(x) \\
    m(x) & -\delta(x) & -\hbar v_F \hat{k}_x & 0 \\
    \delta(x) & m(x) & 0 & \hbar v_F \hat{k}_x \\
  \end{pmatrix},
\end{equation}
where $v_F$ is the Fermi velocity for the helical edge states, $m(x)$ and
$\delta(x)$ represent $x$-dependent coupling between the edge states, and the
basis is ordered as $(R_1,L_1,L_2,R_2)$. The above Hamiltonian manifestly
respects TRI: ${\cal H} = \Theta{\cal H}\Theta^{-1}$. This Hamiltonian can be
derived from microscopic models such as the Bernevig-Hughes-Zhang (BHZ) model
for HgTe/CgTe quantum wells \cite{bernevig06, konig07, konig08, zhou08,
krueckl11}. For simplicity we take $m(x) = m_W \theta(x) \theta(L-x)$ and
$\delta(x) = \delta_W \theta(x) \theta(L-x)$ with $\theta(x)$ the Heaviside step
function and $m_W$ and $\delta_W$ two constants determined by the constriction
separation $W$. In the case of HgTe/CgTe quantum wells, a nonvanishing $\delta$
term owes its existence to the presence of bulk-inversion asymmetry
\cite{konig07, konig08}.

The scattering amplitudes for this constriction, corresponding to $S$ in Eq.
\eqref{eq:s4}, can be easily derived (see supplementary materials):
\begin{align}
  t &= i\cos(\delta_W L/\hbar v_F)/\zeta,\\
  s &= \sin(\delta_W L/\hbar v_F)/\zeta,\\
  r &= m_W\sin(qL)/\hbar v_F q\zeta,
\end{align}
where $q = \sqrt{E^2 - m_W^2}/\hbar v_F$ can be either real or imaginary
depending on the energy $E$, and $\zeta = |(E/\hbar v_F q)\sin(qL) + i\cos(qL)|$
is a normalization factor. Clearly $r$ vanishes when $m_W=0$, which shows
the fact that $m$ couples $R_1$ ($L_1$) to $L_2$ ($R_2$); $s$ vanishes when
$\delta_W=0$, which shows the fact that $\delta$ couples $R_1$ ($L_1$) to
$R_2$ ($L_2$). For low energy ($|E| < m_W$) scattering states, $R/(T+T_s) \simeq
\sinh^2(m_W L/\hbar v_F)$ if $|E|\ll m_W$, and $R/(T+T_s) \simeq (m_W L/\hbar
v_F)^2$ if $|E| \simeq m_W$, meaning that the reflection dominates as long as
$L$ is large compared with $\hbar v_F/m_W$. In this regime, Eq. \eqref{eq:gamma}
reduces to $\gamma \simeq 4n\langle T_s\rangle\langle\sin^2\phi\rangle$.
%For high energy ($|E|\gg m_W$) scattering states, $R/(T+T_s) < (m_W/E)^2 \ll 1$

\begin{figure}
\centering
\includegraphics[width=0.4\textwidth]{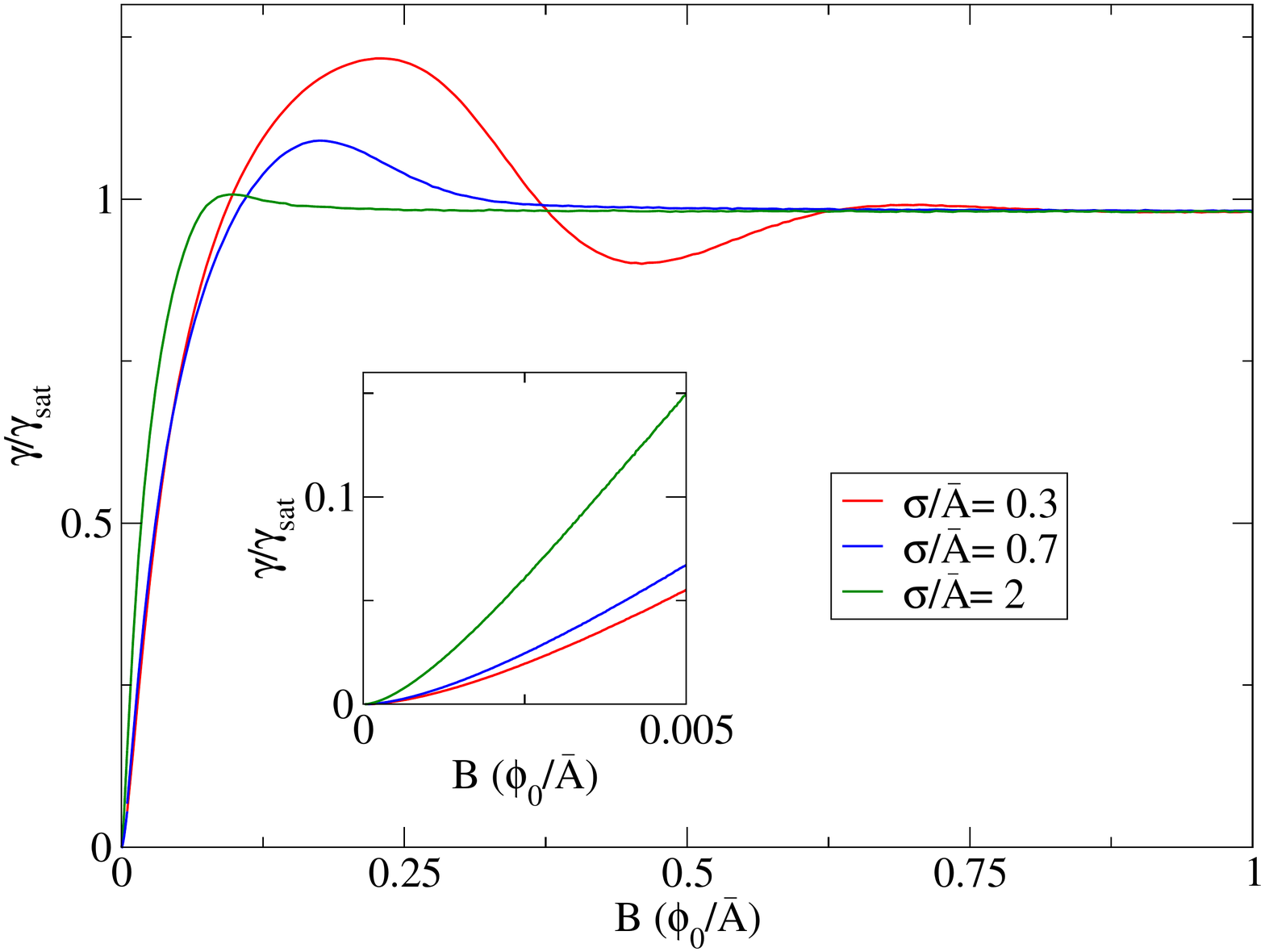}
\caption{Inverse localization length $\gamma$ as a function of magnetic field
$B$, for different distributions of loop area $A$. It shows a $B^2$-increase
in the weak $B$ limit (inset) and a saturation at relatively high field.
In-between, damped oscillations may occur if $\sigma_A/\bar{A} < 1$. The
saturation value $\gamma_\text{sat}$ is sample-dependent but has an order of
magnitude $0.01n$ (with $n$ the linear density of the scatterers) in
our realistic estimations.}
\label{fig:gamma}
\end{figure}

More generally the average with respect to the scattering amplitudes has to be
performed numerically by taking certain distributions of $W$ and $L$ (at a
certain energy $E$). The advantage of this change of variables is that $W$ and
$L$, unlike the scattering amplitudes, are in principle independent to each
other. In total, this leads to three independent geometric variables, $W$, $L$
and $A$, that remain to be averaged on in our final evaluation of $\gamma$ as a
function of magnetic field $B$ (we will not make any assumption on the density
$n$ of scatterers and leave it as a parameter). After carrying out these
averages numerically (see supplementary materials for details), we show the
typical results in Fig. \ref{fig:gamma}. The qualitative behavior of $\gamma(B)$
in Fig. \ref{fig:gamma} is essentially the same as that of $\Gamma(B)$ in Fig.
\ref{fig:Gamma1}, except that $\gamma(B)$ is shown for various ratios
$\sigma_A/\bar{A}$ whereas the scattering amplitudes have been averaged out.
The universal features which we can observe are that $\gamma$ increases as $B^2$ at weak
magnetic field and saturates at sufficiently strong magnetic field. Despite the
fact that the exact value of $\gamma_\text{sat}$ depends on the energy $E$ and
the distributions of $W$ and $L$, the order of magnitude of $\gamma_\text{sat}$
turns out to be consistently $0.01n$ for all cases with realistic considerations
(see supplementary materials). We also observe in the intermediate regime damped
oscillations of $\gamma$ which are pronounced if $\sigma_A/\bar{A}$ is small but
suppressed as long as $\sigma_A/\bar{A}$ is close to or larger than $1$.
We point out here that $\gamma(B)$ has a local minimum/maximum, hence the
localization length has a local maximum/minimum, whenever $B$ is roughly an
integer/half-integer multiple of $\phi_0/2\bar{A}$ -- this is where the TRI is
maximally preserved/broken.

To summarize, we have investigated a simple yet illuminating model
that demonstrates a magnetic-field-dependent localization of the helical
edge states in quantum spin Hall insulators. We have identified universal,
sample-independent features, as well as an interesting but sample-specific
feature in this model. With known parameters for the HgTe/CgTe quantum wells, we
have also estimated quantitatively the localization length. Both the qualitative
and the quantitative results can be examined by experiments.

\acknowledgments
P. D. was supported by the European Marie Curie ITN  NanoCTM and J. L. was
supported by the Swiss National Center of Competence in Research on Quantum Science and Technology.
In addition, we acknowledge the support of the Swiss National Science Foundation. 
The authors would like to thank Alberto Morpurgo and Mathias Albert for inspiring discussions.

\bibliography{biblio_localization}

\begin{widetext}
\section{Supplementary material}
\section{Derivation of Eqs. (19-21)}

In this section we derive the scattering amplitudes, given by Eqs. (19-21) in
the main text, for the constriction described by the Hamiltonian (18) in the
main text. By assuming $m(x) = m_W \theta(x) \theta(L-x)$ and
$\delta(x) = \delta_W \theta(x) \theta(L-x)$ with $\theta(x)$ the Heaviside step
function and $m_W$ and $\delta_W$ two constants, we divide the constriction
into three regions: $x<0$, $0\le x\le L$ and $x>L$. The scattering amplitudes
are obtained by matching energy eigenstate wavefunctions for adjacent regions.

The energy eigenstates in the regions $x<0$ and $x>L$ are trivial:
\begin{align}
  \Psi_a(x<0) =
  \begin{pmatrix}
    a_1 e^{ik_0x} \\
    a_2 e^{-ik_0x} \\
    a_3 e^{-ik_0x} \\
    a_4 e^{ik_0x} \\
  \end{pmatrix}, \qquad
  \Psi_c(x>L) =
  \begin{pmatrix}
    c_1 e^{ik_0x} \\
    c_2 e^{-ik_0x} \\
    c_3 e^{-ik_0x} \\
    c_4 e^{ik_0x} \\
  \end{pmatrix},
\end{align}
where $k_0 = E/\hbar v_F$. In the region $0\le x\le L$, the energy eigenstate
is given by
\begin{equation}
%\begin{split}
\Psi_b(x) =
b_1\left(
\begin{array}{c}
  1             \\ 
- e^{-i\theta}      \\
 e^{-i\theta}   \\
-1
\end{array}
\right)
e^{ik_1x}
+ b_2\left(
\begin{array}{c}
  1             \\ 
e^{i\theta}      \\
e^{i\theta}  \\
 1
\end{array}
\right)
e^{ik_2x}
+ b_3\left(
\begin{array}{c}
  1             \\ 
e^{-i\theta}     \\
e^{-i\theta}  \\
  1
\end{array}
\right)
e^{ik_3x}
+ b_4\left(
\begin{array}{c}
  1             \\ 
- e^{i\theta}      \\
e^{i\theta} \\
-1
\end{array}
\right)
e^{ik_4x},
%\end{split}
\end{equation}
where $\theta = \arccos(E/m_W)$, $k_1 = -k_2 = \delta_W/\hbar v_F  + q$ and $k_3
= -k_4 = -\delta_W/\hbar v_F  + q$ with $q = \sqrt{E^2 - m_W^2}/\hbar v_F =
i(m_W/\hbar v_F)\sin\theta$. Note that both $\theta$ and $q$ can be complex
depending on the energy $E$. Note also that the basis for the above
wavefunctions is ordered as $(R_1,L_1,L_2,R_2)$ (see Fig. 3 of the main text).

To start with we assume that the only incoming state is from channel $R_1$,
that is, $a_1 = 1$ and $a_4 = c_2 = c_3 = 0$. Then we need to match the
wavefunctions such that
\begin{align}
  \Psi_b(x=0) = \Psi_a(x=0) =
  \begin{pmatrix}
    1 \\
    a_2 \\
    a_3 \\
    0 \\
  \end{pmatrix}, \\
  \Psi_b(x=L) = \Psi_c(x=L) =
  \begin{pmatrix}
    c_1 \\
    0 \\
    0 \\
    c_4 \\
  \end{pmatrix} e^{ik_0 L}\,.
\end{align}
The above equations fix the values of $b_i$ $(i=1,2,3,4)$, and hence the values
of $a_2$, $a_3$, $c_1$ and $c_4$. In particular, $a_2$ can be identified as
the backscattering amplitude and we find $a_2=0$ identically; $a_3$ can be
identified as $r$; $c_1$ can be identified as $t$; $c_4$ can be identified as
$s$. We find:
\begin{align}
  t &= i\cos(\delta_W L/\hbar v_F)\sin\theta/\sin(qL-\theta),\\
  s &= \sin(\delta_W L/\hbar v_F)\sin\theta/\sin(qL-\theta),\\
  r &= -i\sin(qL)/\sin(qL-\theta),
\end{align}
which are precisely Eqs. (19-21) in the main text after removing an
unimportant common phase factor. By assuming differently the incoming states,
we can construct the full scattering matrix for the constriction. Both the
unitarity of the scattering matrix and the symmetry relations between the
matrix elements can be easily checked.

\section{Numerical simulations to extract the distributions of scattering
probabilities}

In this section we extract from numerical simulations the distributions of the
scattering probabilities for the constriction illustrated in Fig. 3 of the main
text. In our simulations we employ a six-terminal setup as shown in Fig.
\ref{fig:device}. This setup is equivalent to a Hall-bar setup. We define a
point contact with cosine profiles to simulate the effect of the constriction.
The point contact has two parameters: its length $L$ and its separation $W$.
The depleted regions are defined by a sufficiently high on-site potential
(compared with the band width).

\begin{figure}
\centering
\includegraphics[width=0.45\textwidth]{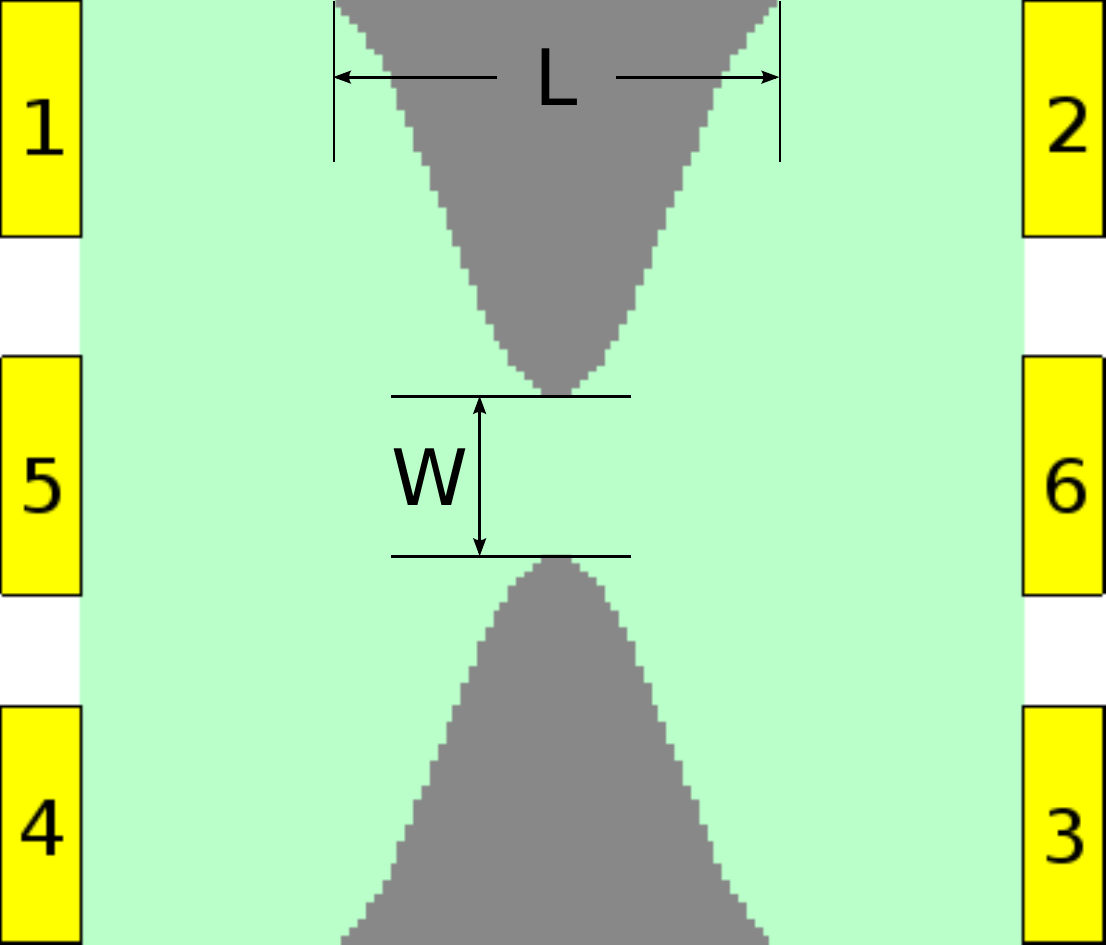}
\caption{Schematic setup for the simulations.}
\label{fig:device}
\end{figure}

The model Hamiltonian for the central region of the setup is the tight-binding
Hamiltonian corresponding to the Bernevig-Hughes-Zhang (BHZ) model
\cite{bernevig06}:
\begin{align}
  & H_{BHZ}(\vec{k}) =
  \begin{pmatrix}
    h(\vec{k}) & -i\Delta\sigma_y \\
    i\Delta\sigma_y & h^*(-\vec{k}) \\
  \end{pmatrix} \\
  \mbox{with}\quad
  & h(\vec{k}) = (C-Dk^2) + Ak_x\sigma_x - Ak_y\sigma_y + (M-Bk^2)\sigma_z
\end{align}
and $\sigma_i$ ($i=x,y,z$) are Pauli matrices. The block-off-diagonal
term, proportional to $\Delta$, is a spin-orbit interaction term due to the bulk
inversion asymmetry in HgTe/CdTe quantum wells \cite{konig07, konig08}. We list
in Table \ref{tab:parameters} the experimentally obtained parameters for the
above Hamiltonian, as well as the lattice spacing $a$ adopted to discretize this
Hamiltonian. We will measure length in units of $a$ hereafter. With these
parameters we estimate the Fermi velocity for the helical edge states to be
$v_F\simeq 3.8\times10^5$ m/s.
\begin{table}
  \begin{tabular}{c|c|c|c|c|c|c}
    $A$ (meV$\cdot$nm) & $B$ (meV$\cdot$nm$^2$) & $C$ (meV) & $D$
(meV$\cdot$nm$^2$)& $M$ (meV) & $\Delta$ (meV) & $a$ (nm) \\
    \hline
    364.5 & -686 & -7.5 & -512 & -10 & 1.6 & 5 \\
  \end{tabular} 
\caption{Parameters for the simulations. Note that the parameter $C$, which
corresponds to an overall constant energy shift, takes the value so that the
edge bands cross at $E=0$.}\label{tab:parameters}
\end{table}

The scattering probabilities for the constriction that are needed in the main
paper can be identified in the current setup as follows:
\begin{align}
  T = T_{21},\; T_s = T_{31},\; R = T_{41},
\end{align}
where $T_{ji}$ is the transmission probability from contact $i$ to contact $j$.
These transmission probabilities can be calculated by using the standard
Green's function technique \cite{fisher_relation_1981}. We also check the sum
rule $T_{sum} = T_{21} + T_{31} + T_{41} = 1$ to ensure the validity of our
results.

For the distributions of the scattering probabilities we need to assume
reasonable distributions for $L$ and $W$. Without sufficient knowledge from
experiments we make the following assumptions: $L$ is uniformly
distributed in the range $(0,2l_{SO})$ with the spin-orbit length
$l_{SO} = \hbar v_F/\Delta \simeq 30a$; $W$ is uniformly distributed in the
range $(0,2\xi)$ with $\xi$ the penetration depth of the edge states. $\xi$ is
energy dependent and its order of magnitude is given by $\hbar v_F/M \simeq
5a$. We will fix the energy at $E=0.5$ meV for the results presented below,
where $\xi \simeq 10a$. We have checked the robustness of our results with other
values of energy and/or other reasonable distributions (e.g. Gaussian
distribution) of $L$ and $W$.

\begin{figure}
\centering
\includegraphics[width=0.9\textwidth]{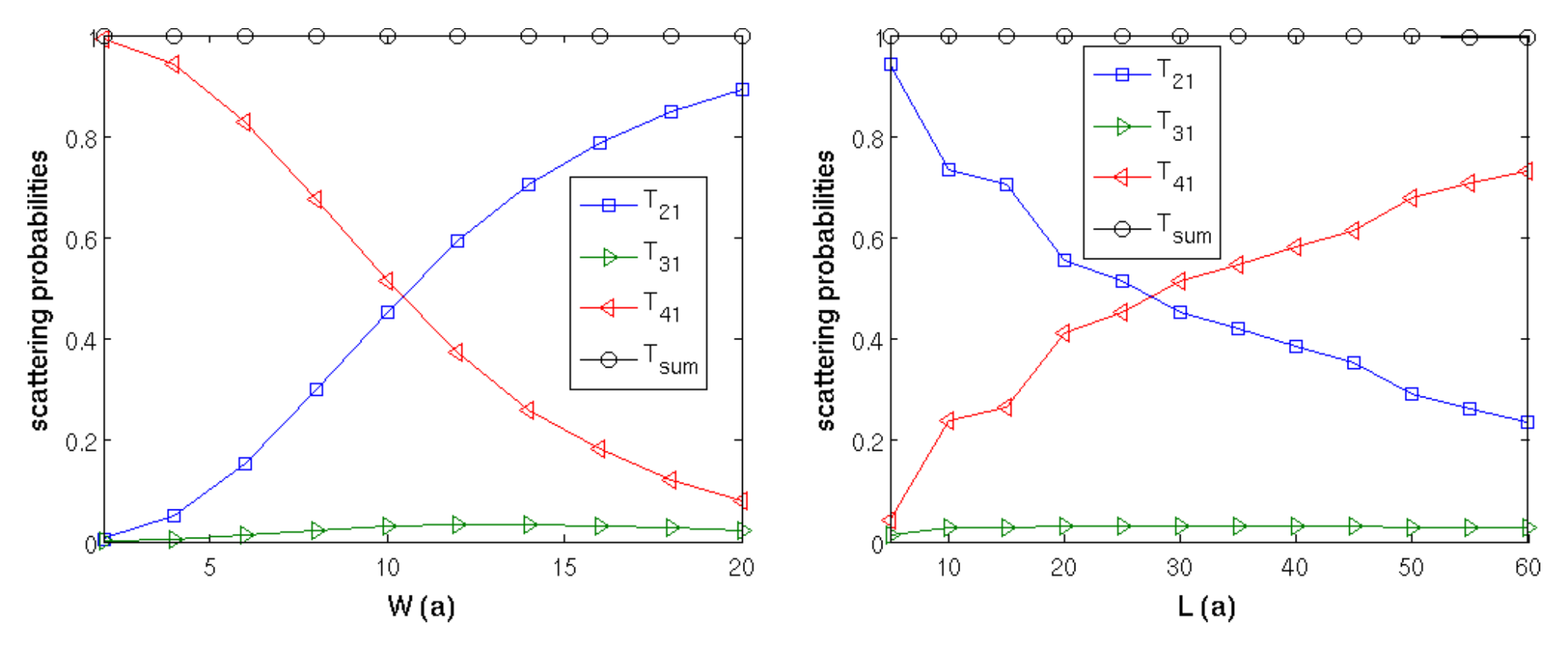}
\caption{Typical dependences of scattering probabilities on $W$ and $L$. For
the left panel, $L=30a$; for the right panel, $W=10a$.}
\label{fig:wl}
\end{figure}

\begin{figure}
\centering
\includegraphics[width=0.55\textwidth]{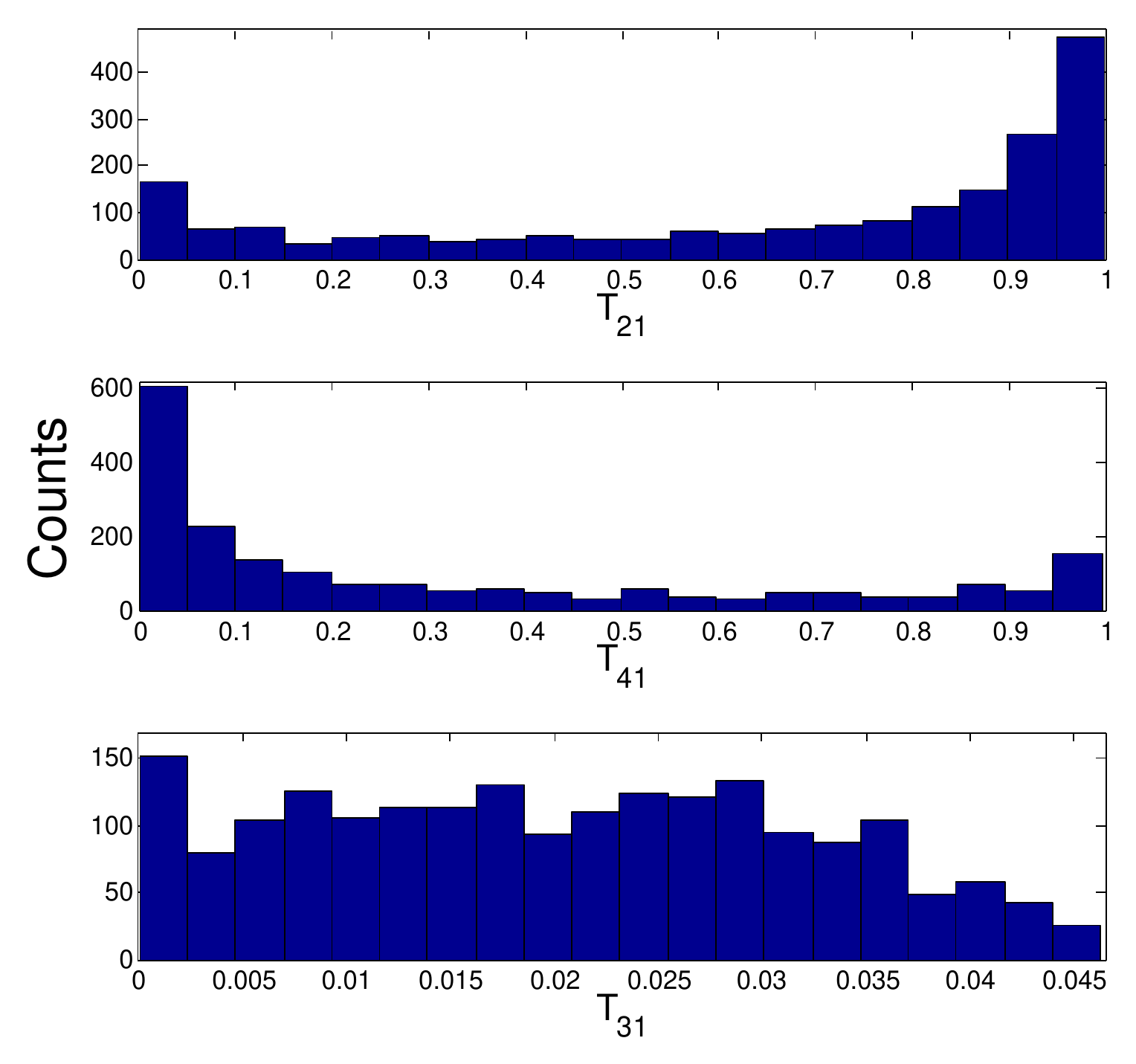}
\caption{Histograms showing the distributions of the scattering probabilities.}
\label{fig:hist}
\end{figure}

As results of our simulation, we show the typical dependences of the scattering
probabilities on $L$ and $W$ in Fig. \ref{fig:wl}, and the histograms for
the distributions of the scattering probabilities in Fig. \ref{fig:hist}. To
obtain our final result presented in Fig. 4 of the main text, we simply
generate samples with randomly chosen $L$ and $W$ according to our
assumptions; no accurate distribution functions for the scattering
probabilities are actually needed for our purpose.

\end{widetext}

\end{document}